\documentclass[useAMS,singlespacing]{mnras}
\usepackage{graphicx,amssymb}

\usepackage{newtxtext,newtxmath}
\usepackage{xcolor}
\usepackage{ulem}

\title[Echoes of the past: ultra-high energy cosmic rays accelerated by radio galaxies, scattered by starburst galaxies]
{Echoes of the past: ultra-high energy cosmic rays accelerated by radio galaxies, scattered by starburst galaxies}
\author[AR Bell, JH Matthews]
{AR Bell$^{1,2}$\thanks{E-mail:Tony.Bell@physics.ox.ac.uk},
JH Matthews$^{3}$ \\
$^{1}$University of Oxford, Clarendon Laboratory, Parks Road, 
Oxford OX1 3PU, UK\\
$^{2}$Central Laser Facility, STFC Rutherford Appleton Laboratory, Harwell Oxford, Oxfordshire OX11 OQX, UK\\
$^{3}$Institute of Astronomy, University of Cambridge, Madingley Road, Cambridge, CB3 0HA, UK
\\
}
\begin{document}
\date{Accepted 2022 January 4. Received 2021 December 18; in original form 2021 August 20.} 

\pagerange{\pageref{firstpage}--\pageref{lastpage}} \pubyear{2022} 
\maketitle
\label{firstpage}
\begin{abstract}
We explore the possibility that the hotspot of ultrahigh energy cosmic rays (UHECR) detected by the Telescope Array (TA) from the approximate direction of M82 and the M81 group of galaxies might be the echo of UHECR emitted by Centaurus A in an earlier more powerful phase.
Echoes from other starburst galaxies or groups of galaxies may contribute to the UHECR flux at the Earth.
{
We use an illustrative Monte Carlo model of mono-energetic UHECR transport by small angle scattering to generate synthetic sky maps.
The model informs a discussion of overall energetics and time and distance scales.
We find a viable echo model for the observed UHECR hotspots if the UHECR luminosity of Centaurus A 20 Myr ago was
200 times its present luminosity and if the ordered magnetic field exceeds 10-20nG out to a distance of 400-800kpc in the circumgalactic medium of M82 and other starburst galaxies.
}
\end{abstract}
\section{ INTRODUCTION}
The origin of ultrahigh energy cosmic rays (UHECR) remains one of the most challenging unknowns in astrophysics.
Directional data from  the Pierre Auger Observatory (PAO) in the southern hemisphere and the Telescope Array (TA) in the northern hemisphere give hope that we are close to identifying their source.
The Hillas (1984) parameter $E_{\rm max}=ZuBR$ (in SI units with $E_{\rm max}$ in eV) constrains the maximum attainable cosmic ray (CR) energy $E_{\rm max}$, where $u$, $B$ and $R$ are the characteristic velocity, magnetic field and spatial scale and $Ze$ is the UHECR charge.
As pointed out by many authors (Lovelace 1976; Waxman 1995; Waxman 2001; Blandford 
2000), the Hillas parameter is related to the magnetic power $P_{\rm mag}$ of the source:
$ZuBR=Z(u \mu_0 P_{\rm mag})^{1/2}$ where $P_{\rm mag}=uR^2B^2/\mu_0$.
This relation can be inverted to give a lower limit on the magnetic power, $P_{\rm mag}>E_{\rm max}^2 /(Z^2 u \mu _0)$.
The total source power must exceed this because (i) the magnetic energy is only one part of the total energy, (ii) the power channeled through the accelerating structure, probably but not necessarily a shock,
is a fraction $\xi$ of the total source power such that $P_{\rm total}=\xi ^{-1} P_{\rm mag}$.
Taking characteristic values suitable for radio galaxies, the total power can be estimated:
$$
P_{\rm total} \sim 2 \times 10^{44} Z_6^{-2} E_{100}^2 (u/0.3c)^{-1}\xi_{0.1}^{-1} \ {\rm erg\ s}^{-1}
\eqno{(1)}
$$
where $Z=6Z_6$, $E_{\rm max}=100 E_{100}\ {\rm EeV}$, $\xi=0.1 \xi_{0.1}$.
Taken together, the Hillas condition and the related power requirement restrict the range of possible sources of UHECR.
The stringent power requirement points to radio galaxies (RG) or gamma-ray bursts (GRB, {e.g. Vietri 1995}) as possible contenders.

GRB easily reach the required power due to the short time in which energy is released, but there are many issues
that question the viability of GRB as an explanation of UHECR origins. 
These issues relate to 
limits placed by the expected associated neutrino fluxes, radiation losses during UHECR acceleration and escape, host galaxy metallicity, UHECR composition
and the ability of GRB to provide the entire UHECR flux
(Hillas 1984, {Guetta et al 2004, Becker et al 2006}, Waxman \& Bahcall 1997, 1998, Aharonian et al 2002,
Ptitsyna \& Troitsky 2010, Dermer 2011, Baerwald et al 2015, 
Anchordoqui 2018, Zhang et al 2018, Alves Batista et al 2019, Boncioli 2019, Samuelsson et al 2019; Heinze et al 2020, Rudolph et al 2021).
An additional difficulty with GRB as a producer of UHECR is that relativistic shocks are limited as accelerators to ultra-high energies (Lemoine \& Pelletier 2010, Reville \& Bell 2014, Bell et al 2018).

Here we focus on RG as a natural source of UHECR because they are large, powerful, and long-lived.  
Also, radiative losses are not a strong limitation if acceleration takes place
in the lobes where magnetic and radiation energy densities are relatively small.
RG display features such as shocks, shear flows, and magnetic reconnection that are known to be suitable for CR acceleration.

Cygnus A is a clear candidate for UHECR production since it far exceeds the power requirement (Eichmann et al, 2018),
but its distance of 237Mpc (Persic \& Rephaeli 2020) counts against it (Eichmann 2019a) because 
(i) UHECR from Cygnus A are unlikely to arrive at the Earth anisotropically without scattering since a magnetic field of only $0.03{\rm nG}$ is needed
to make their Larmor radius smaller than the distance to Cygnus A
(ii) even without scattering, the UHECR travel time from Cygnus A is many times the loss time due to the
Greisen–Zatsepin–Kuzmin (GZK; Greisen
1966; Zatsepin \& Kuz’min 1966) effect and photodisintegration
(e.g. Stecker \& Salamon 1999, Hooper, Sarkar \& Taylor 2007).

In fact, powerful FRII galaxies are rare within the local GZK volume (Massaglia 2007).
FRI radio galaxies are more populous, and radio galaxies at the FRI/FRII boundary have powers close to $P_{\rm total}$ as given in equation (1).
The most obvious FRI candidate is Centaurus A (Cen A, {e.g. Biermann \& Souza 2012}) which is nearby and estimated to have a cavity power of
$10^{44}{\rm erg \ s}^{-1}$ (Hardcastle et al 2009, O'Sullivan et al 2009, Cavagnolo et al 2010, van Velzen et al 2012, Matthews et al 2018).
The case for Cen A is strengthened by the coincidence of Cen A with the brightest UHECR hotspot detected by PAO (Aab et al 2018) and the apparent need for local UHECR sources (Guedes Lang et al 2020).
PAO also detects a weaker hotspot with weaker statistical significance in the approximate direction of the radio galaxy Fornax A
which has similar characteristics to Cen A
(Matthews et al 2018), including evidence of variability (Mackie \& Fabbiano 1998, Maccagni et al 2020).

The arrival directions of UHECRs favour starburst galaxies or AGN as likely sources.
Aab et al (2018) and
Bister et al (2021) find a statistically significant correlation of PAO arrival directions with active galaxies, but find an even stronger correlation with starburst galaxies.
This starburst correlation is broadly supported by van Vliet et al (2021), who find the Auger anisotropy measurement is consistent with the local star-forming galaxy density for nG level extragalactic magnetic fields.
Further evidence pointing toward starburst galaxies and away from radio galaxies is the lack of a powerful radio galaxy in the direction of the UHECR hotspot
detected by TA in the northern hemisphere.
A note of caution is sounded by the analysis of Abassi et al (2018) 
which does not find a a statistically significant association of TA arrival directions with starburst galaxies.
Nevertheless, the TA hotspot lies in the approximate direction of the nearby strong starburst galaxy M82.

More data is needed but, as observations stand, starburst galaxies appear to be a strong candidate for the origins of UHECR.
However, their powers reach typically only $10^{42}{\rm erg \ s}^{-1}$ 
and the characteristic wind velocity in a starburst region is $\sim c/300$, 
(Heckman, Armus \& Miley 1990; Anchordoqui 2017; Romero,
Muller \& Roth 2018, Matthews et al 2020), 
so they fail to reach the power condition (equation 1) for a source of UHECR.  
We therefore proceed by setting aside starburst galaxies as a direct source of UHECR despite their statistical correlation with UHECR arrival directions.
The possibility remains that UHECR are produced by GRB  associated with starburst galaxies.

Here we explore the possibility that starburst galaxies or associated galaxy groups are instead reflectors of UHECR without themselves being a source of UHECR.  
We propose that the environs of starburst galaxies are regions of enhanced scattering that diffusively reflect UHECR back towards the Earth after their initial acceleration by Cen A.

\section{ THE MODEL}

We initially develop our model by focussing on diffusive reflection from  the prominent starburst galaxy M82.
In a later section we expand the model to include diffusive reflections from other nearby starburst galaxies.

We suggest that the TA hotspot is associated with the M81 group of galaxies and in particular with M82 which,
being 39kpc from M81, lies close to the centre of the group.
These two galaxies and a smaller starburst galaxy NGC3077 at a distance of 49kpc from M81 form a mutually interacting triad 
with a history of mass transfer (de Mello et al 2008, Smercina et al 2020).
They are neighbours of the Milky Way at a distance of 3.7Mpc from the Earth,
which is similar to the distance of Cen A from the Earth.
The M81 group of 30-35 galaxies has an outer radius of about 800-1000kpc (Karachentsev 2005, Karachentsev \& Kashibadze 2006).
Based on the Larmor radius of a 50EeV UHECR with $Z=6$, an ordered magnetic field of 10nG over 800kpc
would be sufficient for the galaxy group to be an obstacle to UHECR propagation.
Such a  magnetic field scatters UHECR away from their previous direction of propagation
and generates a {broadly spread} flux in the reverse direction {(diffusive reflection) as seen in figure 3}.
The interaction with the magnetic field is inevitably complex and we model it reductively as small angle scattering.

M82 has a mass of $\sim 10^{10}{\rm M_\odot}$ (Greco et al 2012).  M81 has a mass twice that of M82 (Karachentsev \& Kashibadze 2006).
Wilde et al (2021) have surveyed HI column densities of 572 galaxies and set a criterion for measuring the radius of the associated circumgalactic medium (CGM).
For galaxies with masses similar to M81 and M82 they find a characteristic radius of 350kpc with the CGM extending out to 1Mpc in some cases.
Tumlinson et al (2011) and Borthakur et al (2013) find that the strong outflows of starburst galaxies affect the CGM out to larger radii
 than other galaxies,
so it is reasonable to suppose that the radius of the M82 CGM is 
at the larger end of the range measured by Wilde et al (2021).
According to the results of Wilde et al, although NGC3077 is less massive, its halo may not be much smaller than those of 
M81 and M82.
Hence, the combined outflows from these galaxies can be expected to act as a significant component of the group environment out to 100s of kpc.

Starburst galaxies can generate a large magnetic field of $100-300\mu{\rm G}$, in the central kpc
 (Domingo-Santamari \& Torres 2005, de Cea del Pozo, Torres \& Marrero 2009, Anchordoqui 2018)
which may be carried 100s kpc into the surrounding medium by the superwind
{
which expands at a velocity exceeding the escape velocity of the galaxy
(Irwin et al 2019).
From radio observations of M82 Adebahr et al  (2013, 2017)  conclude that the magnetic field is frozen into the outflowing wind
and driven out of the galaxy kinetically to feed magnetic field into the surrounding medium.
}

A tidal interaction between M81 and M82 a few hundred Myr ago (de Mello et al 2008) is thought to have triggered the present starburst activity in M82.
Lopez-Rodriguez et al (2021) measure a median large scale magnetic field of $305\mu {\rm G}$ in the central 900 pc of the M82 starburst.
They estimate a field of $15\mu {\rm G}$ at a distance of 6.6kpc from the centre of the starburst.
$15\mu{\rm G}$ is sufficient to form a magnetic obstacle to 50EeV UHECR with $Z=6$
since their Larmor radius is then 0.6kpc.
Lopez-Rodriguez et al (2021) find that the field lines are  `open' and feed magnetic flux into the surrounding medium.

{
The magnitude and structure of the magnetic field can be deduced in the central few kpc of M82
from radio obervations (Adebahr et al 2013, 2017), and analytical models can suggest
possible configurations (Ferrière \& Terral 2014), but the magnetic field at 100s kpc 
is less easily assessed.
Nevertheless, the principles of frozen-in flow in a spherical wind offer some guidance.
Whatever the structure of the inner magnetic field, tangential stretching of the field during outflow
enhances the tangential components relative to the radial component.
Shear flow due to turbulence or non-uniform expansion could rotate the field lines to enhance the radial
components without reducing the tangential components which may also grow due to turbulence.
If, for simplicity, we assume that }
the magnetic field is frozen-in to a constant velocity uniform spherical ($r,\theta,\phi$) flow,
the radial component $B_r$ is proportional to $r^{-2}$
while the tangential components, $B_\theta$ \& $B_\phi$ decrease more slowly in proportion to $1/r$,
giving 
$$
B(r)=B_0\left ( \epsilon _\perp  \frac {r_0^2}{r^2} +(1-\epsilon_\perp) \frac {r_0^4}{r^4}\right )^{1/2}
\eqno{(2)}
$$
where $B_0$ is the magnetic field at radius $r_0$, and $\epsilon _\perp$ is the fraction of the magnetic energy in the tangential field at $r_0$.
If the field is randomly oriented ($\epsilon _\perp =2/3$) and $B_0=15\mu {\rm G}$ at $r=6.6{\rm kpc}$,
then at 800kpc, $B= 100 {\rm nG}$.
On the other hand, if the field were purely radial to the extent that $\epsilon _\perp \ll 7 \times 10^{-5}$,
which is ruled out by $\nabla \cdot {\bf B}=0$ except in extreme geometries, then
$B=1{\rm nG}$ at 800kpc.

Samui et al (2018) model the outflow from a normal galaxy with a mass of $10^{11}{\rm M_\odot}$.
In their `conservative model', they find that a magnetic field of 50nG at 10kpc falls to $1.5{\rm nG}$ at 300kpc.
Their fields are much smaller than fields in starburst galaxies, but their dependence on radius is approximately $B\propto r^{-1}$
as appropriate for frozen-in constant velocity expansion.
Their `optimistic' model allows for turbulent amplification of magnetic field (Bertone et al 2006, Donnert et al 2018) 
and points to a magnetic field that is larger by an order of magnitude.
Indeed, the mutual interaction of the outflows from M81, M82 and NGC3077 increases the likely incidence of
sheared flow and turbulence which lead to magnetic field amplification.
The magnetic field at large radius is also larger if the flow velocity slows down and compresses the tangential field.

We need to ensure that the magnetic energy density is not excessive.
Strickland \& Stevens (2000) and Strickland \& Heckman (2009) estimate the total output mechanical power of the M82 starburst region as ${\dot E}=3\times 10^{42}{\rm erg \ s}^{-1}$.
In comparison, the total magnetic energy of a 800kpc radius sphere of 10 nG magnetic field is $E_{\rm mag}=2.5\times 10^{56}{\rm erg}$.
The ratio $E_{\rm mag}/\dot{E}$ is 3Myr suggesting that the energy requirement of a 10nG magnetic field is 
not excessive if injected over 100s Myr or more during which M81 and M82 have been interacting.

Overall, the structure and strength of the magnetic field environment is clearly uncertain; 
however, the existence of a 10nG field throughout the halo  appears reasonable. 
This is especially true when considering possible magnetic field amplification and/or pollution from starburst-driven outflows.

We therefore build a model of UHECR transport on the basis that UHECR are scattered by a combination of M82, M81, NGC3077 and their mutual interacting outflows.
In our model, UHECR are produced by Cen A 
{
as discussed below.
}
Some UHECR propagate directly to the Earth.
Other UHECR arrive later via M82 and the M81 galaxy group where they are scattered and diffusively reflected.
We initially assume that the radius of the halo of M82 is 800kpc.
In section 7 we present results for a smaller halo radius of 400kpc.

The spatial extent of the TA hotspot is uncertain.
di Matteo et al (2019) identify the hotspot by drawing circles of radius $15^\circ$ or $20^\circ$ around its central maximum.
 $15^\circ$  corresponds to $1{\rm Mpc}$ at the distance of M82.
In their analysis Kawata et al (2015) use a radius of $20^\circ$.
The  $15-20^\circ$ hotspot is embedded in a larger structure and the limited statistics of UHECR arrivals are insufficient
to realistically make a statement on the shape and size of the hotspot and the surrounding structure.
It is possible as more data are accumulated that the word 'hotspot', with its connotations of a single point source,
may become an inapproriate designation of the UHECR excess in the direction of M81 and M82.
A further complication is that the 
size of the hotspot on the sky is due to a combination of  the size of the source and UHECR scattering during propagation to the Earth.

Cen A, M82 and the Earth form an approximately isosceles triangle with M82 and Cen A both about 3.7Mpc from the Earth as shown in figure 1. 
The corresponding distance between Cen A and M82 is 6.4Mpc.
The UHECR travel time from Cen A to the Earth via M82 is longer by 21Myr,
which means that the UHECR luminosity of the M81 group reflects the luminosity of Cen A 33Myr ago as it would have appeared at the Earth 21Myr ago. 


As discussed by Matthews \& Taylor (2021), the time variation of UHECR production by radio galaxies depends on variation in accretion onto the central black hole, 
the hydrodynamic response of the jets and lobes, the sensitivity of the acceleration process to the presence of shocks and turbulence, and the residence time of UHECR in the lobes. 
Matthews \& Taylor (2021) consider a jet power that varies according to a flicker noise power spectrum with a log-normal distribution of jet powers, 
motivated by the chaotic cold accretion model of Gaspari et al. (2017) and numerical simulations of AGN fuelling (Yang \& Reynolds 2016; Beckmann et al 2019); 
these studies suggest that variation in jet power by orders of magnitude on a 20Myr timescale is reasonable even before considering the individual history of Cen A. 
UHECR production can be especially spasmodic due to the need to exceed the power threshold. 
Conditions for UHECR acceleration may exist only during peak periods of activity with CR failing to reach UHECR energies during quiescent periods. 
In our picture, Cen A is in a relatively inactive state at the present time with weak UHECR production compared with 20~Myr ago when it was more active. We note that this timescale is comparable to the spectral age of the synchrotron electrons in the giant lobes, estimated to be $\sim25$~Myr by Hardcastle et al (2009).

From repeated application of the inverse square law for a halo radius of 800kpc, the UHECR flux arriving at the Earth via M82 
is $\sim 100 \times$ smaller than the flux arriving directly from Cen A if the same number of UHECR are released 
in the directions of M82 and the Earth.
This is confirmed by figure 2 below where we show the results of a Monte Carlo calculation.
Hence, for our model to work, the UHECR output of Cen A must have been considerably greater 20Myr ago
as discussed quantitatively in Sections 4 and 7 below.

It is not unreasonable that the UHECR lumnosity of Cen A over longer timescales of 10sMyr should 
be greater than its present luminosity.
We have previously suggested (Matthews et al, 2018, 2019) that the giant inflated lobes of Cen A act as UHECR reservoirs
and that the present flux of UHECR direct from Cen A is the leakage from the lobes of UHECR accelerated during past
active episodes.
This would explain how UHECR can now be arriving from Cen A when the present jet power of $\sim 10^{43}{\rm erg \ s}^{-1}$
(Russell et al 2013, Wykes et al 2013, Eilek 2014, Matthews et al 2018)
falls short of that required by equation (1).
An additional indication of past enhanced activity is that
the size and energy content of the large radio lobes is large in comparison with the 
present weak radio jets that extend only a small distance into the lobes.

Further indications that the UHECR power of Cen A might have been greater in the past are 
(i) the present UHECR (energy greater than 55EeV) power of Cen A, estimated to be $2 \times 10^{39}{\rm erg \ s}^{-1}$ (Joshi et al 2018), is very much smaller than 
the Eddington luminosity of $7\times 10^{45}{\rm erg \ s}^{-1}$ 
corresponding to its central black hole mass of $5.5 \times 10^7$ solar masses (Capellari et al 2009)
(ii) the large energy content, $10^{59}-10^{60}{\rm erg}$ of the lobes (Wykes et al. 2013; Eilek 2014), suggests a larger energy input in the past
(iii)  previous enhanced activity is consistent with a history 
 of galaxy mergers  in the past 1-2 Gyr  (Eilek 2014, Wang et al 2020).

In the next section we explore this model further with the use of Monte Carlo calculation of a single burst of UHECR emitted by Cen A,
encountering a sphere of enhanced scattering representing the M81 group, and propagating from there to a detector at the Earth.
The actual history of Cen A is probaby more complicated with multiple powerful episodes over much longer periods of time.  
This is beyond the scope of the present paper.

\section{ AN ILLUSTRATIVE MONTE CARLO CALCULATION}
We use a Monte Carlo model of small angle UHECR scattering to examine the possibility that the TA hotspot is due
to enhanced scattering by the M81 group of galaxies and principally by the halo of M82.

{
Given the uncertainties in the magnetic field both within the halo of M82 and in the intergalactic
medium between Cen A, M82 and the Earth,
we do no attempt to track the propagation of UHECR through specified magnetic fields.
Instead we construct a model based on small-angle scattering by disordered magnetic fields.
This may be realistic in the intergalactic medium if the magnetic field is structured on scales smaller than the distance between Cen A and M82, which is in turn smaller than the UHECR Larmor radius. 
The assumption of small angle scattering may well be incorrect in the halo of M82.
Nevertheless, the model serves its essential purpose
of making the halo a barrier to UHECR propagation and
scattering UHECR back into the intergalactic medium with a broad spread of angles.
A breakdown of the small angle scattering assumption inside the halo would produce a hotspot with angular structure,
and this would be consistent with UHECR observations.
Future work based on a knowledge of magnetic field structures may improve upon our model.
As discussed below, our model treats the UHECR as mono-energetic with envisaged energies of 50-60EeV since
these contribute most to the PAO and TA hotspots.
}
{
The UHECR composition is uncertain.
Results from PAO indicate an intermediate nuclear composition 
(Todero Peixoto et al 2019) and we envisage $Z\sim 6$ as characteristic.
The crucial parameter in our model is the UHECR scattering rate.  
We choose a scattering rate that produces hotspots on the sky with a size that is consistent
with observations (see figures 6 and 8).
}

The computer code uses the Boris rotation algorithm (eg Birdsall \& Langdon 1985)  to randomly rotate the direction of propagation by $\Delta \theta = 1.43^\circ$ 
at time intervals of $\Delta t_{\rm Myr}= 0.063{\rm Myr}$.
For small deflections, the RMS deflection after $t_{\rm Myr}$ is $\theta =(t_{\rm Myr}/\Delta t _{\rm Myr})^{1/2} \Delta \theta $.
This gives a deflection of $20^\circ$ after 12Myr
during which time  it travels a distance of 3.7Mpc.
This is applied in the intergalactic medium outside the halo.
{
The scattering rate is chosen to approximately match the angular size of the PAO hotspot coinciding with Cen A.
The scattering rate is a function of UHECR charge, magnitude of magnetic field and coherence lengths.  
For example, the scattering rate may be the result of relatively strong magnetic field varying randomly over scales very much
smaller than the UHECR Larmor radius.  Or, alternatively it may result from relatively weak fields varying over distances
that are  a larger fraction of the Larmor radius.
Our calculation does not distinguish between these different possibilities.
For comparison with our scattering rate,
a UHECR with energy 55EeV and charge $Z=6$ is deflected through an angle of $6^\circ$ 
by a uniform 1nG field while
travelling a distance of 1Mpc.
}
{
Inside the halo we increase the scattering rate and reduce the timestep $\Delta t _{\rm Myr}$ as described below.
}

Given the uncertainty in crucial parameters and in the scattering process, the model is illustrative rather than definitive.
Future progress in understanding the structure of intergalatic magnetic fields may facilitate improved modelling
using sophisticated UHECR propagation codes
such as CRPropa ({Alves Batista et al 2016, Merten et al 2017}).
In this section we proceed by assuming a halo radius of 800kpc.
Results for the smaller 400kpc halo are presented below in Section 7.
Many detailed variants of our basic model might be considered.

Our picture is that the lobes of Cen A act as a reservoir of UHECR (Matthews et al 2018, 2019, 2020).
UHECR are injected into the reservoir over a period of $\sim 4 {\rm Myr}$ and varying in time as a Gaussian with $\sigma = 2 {\rm Myr}$.
UHECR then leak out of the reservoir with an exponential decay time of $3 {\rm Myr}$. 
The decay time is very uncertain since it depends on UHECR transport in magnetic field with an unknown structure.
Our decay time of $3{\rm Myr}$  is
 chosen to reproduce the observed ratio of the Cen A and M82 hotspots as shown in figure 2.
It is consistent with, if somewhat shorter than, the estimates made by Matthews \& Taylor (2021).
In the calculation, the time $t=0$ corresponds to the peak in UHECR production by Cen A.
The UHECR arriving at the Earth via M82 are delayed $20{\rm Myr}$ relative to those arriving by direct propagation from Cen A.
We ignore GZK and other losses since they occur on a longer timescale.

We treat the UHECR as mono-energetic.
Each UHECR is given the same spatially uniform scattering rate outside the halo of M82.
The mono-energetic assumption is a good representation because the UHECR energy spectrum is steep and dominated by UHECR with energies close to the lower energy cut-off,
{
although this justification is complicated by composition as mentioned above.
}
Some UHECR arrive directly at the Earth.  Other UHECR are scattered by M82 halo to arrive at the Earth from the direction of M82.
We adopt the geometry shown in figure 1 where the distance of both Cen A and M82 from the Earth is 3.7Mpc,
and the distance between Cen A and M82 is 6.4Mpc.
The spherical halo of enhanced scattering around M82 has a radius of 800kpc.

\begin{figure}
\includegraphics[angle=0,width=5cm]{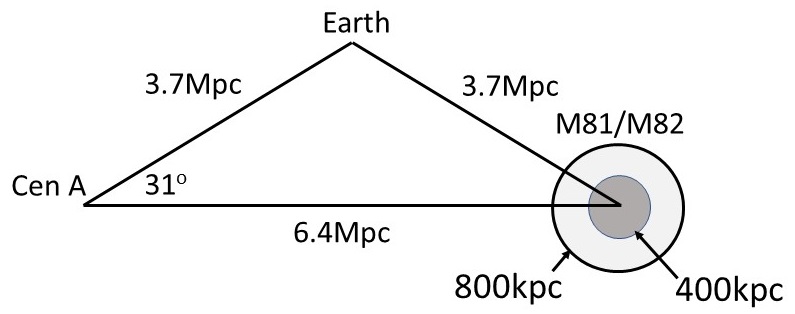}
\centering
\caption{
Geometry of Centaurus A, M81/M82 and the Earth. 
}
\label{fig:figure1}
\end{figure}

\begin{figure}
\includegraphics[angle=0,width=8.5cm]{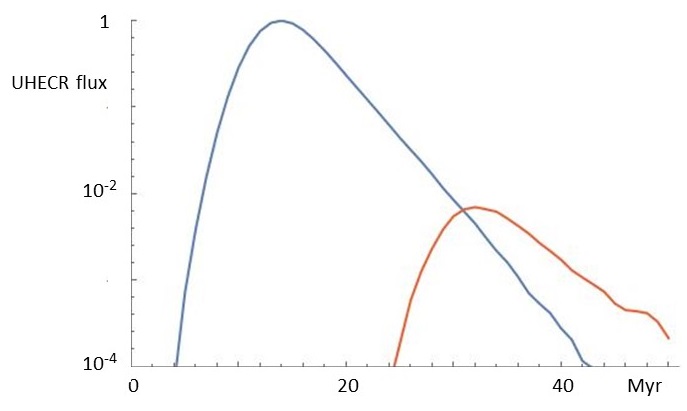}
\centering
\caption{
Time evolution of the logarithmic UHECR intensity at the Earth (i)  UHECR arriving directly from Centaurus A without passing through
the halo of M82 (blue line) (ii) UHECR arriving at the Earth having passed through the halo of M82 (brown line).
}
\label{fig:figure1}
\end{figure}
A sphere of enhanced scattering represents the halo of M82.
The sphere is placed on the horizontal $z$ axis, thus imposing cylindrical $(r,z)$ symmetry on the calculation.
The enhanced scattering rate in the halo is chosen to be forty times that in the intergalactic medium.
This is sufficient to make the scattering sphere optically thick to UHECR.
The scattering rate determines how far the UHECR penetrate into the sphere before reflection back to the surface.
Its value makes little difference to the results provided optical thickness is maintained.
We add a notional detector at the position of the Earth and analyse how the UHECR 
number density and anisotropy evolve in time.

The UHECR density  in $(r,z)$ is plotted in figure 3.
The shell of UHECR expanding away from Cen A can be clearly seen.
The number of UHECR reflected by M82 is smaller and is not easily seen in the top row of figure 3
apart from those within the M82 halo.
The bottom row of figure 3 plots only those UHECR that have (i) passed through the halo of M82, and (ii) are not at present within the halo.
The intensity of reflected UHECR is stronger in the direction back towards Cen A with a broad spread in angle
that encompasses the Earth.

\begin{figure}
\includegraphics[angle=0,width=8.5cm]{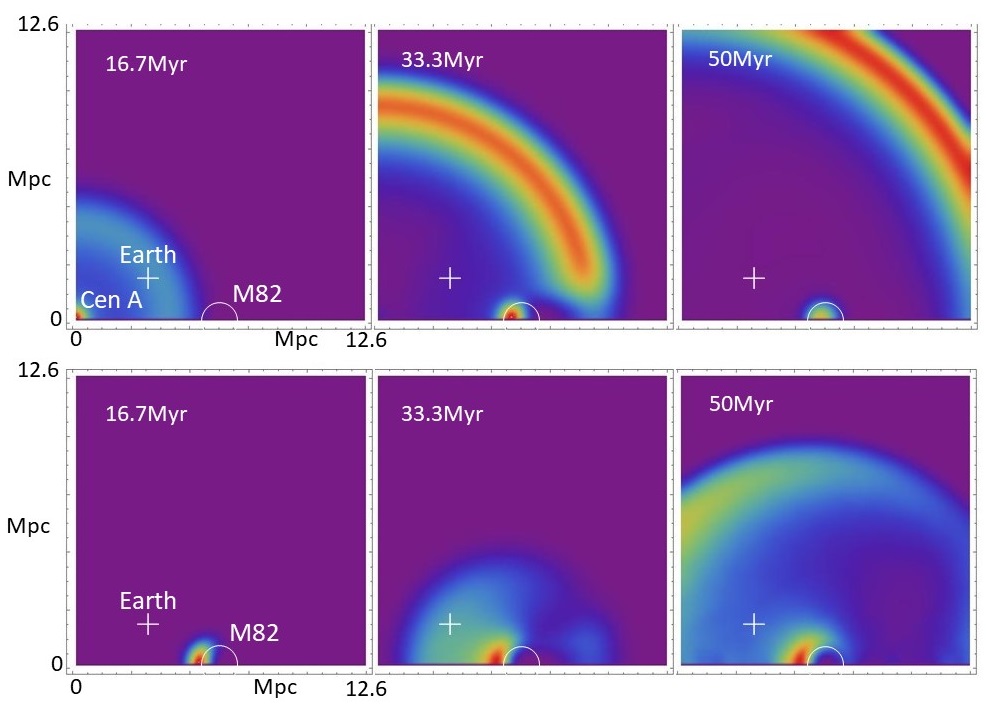}
\centering
\caption{
The top row gives the density plotted in $(r,z)$ of all UHECR at times 16.7, 33.3 and 50Myr.
The bottom row is the density of only UHECR that have at some time in their prior history passed
through the halo of M82 but are not at present within the halo.
In each plot the colour scales are scaled to the maximum UHECR density.  The absolute scaling can be deduced from Figure 2.
}
\label{fig:figure1}
\end{figure}
Figure 4 plots the angle of the mean streaming velocity of UHECR arriving at the Earth.
The angle is defined relative to the $z$ axis connecting Cen A to M82.
As expected, the angle initially corresponds to arrival from the direction of Cen A and then swings round to
point at M82 when the reflected UHECR arrive.
\begin{figure}
\includegraphics[angle=0,width=8.5cm]{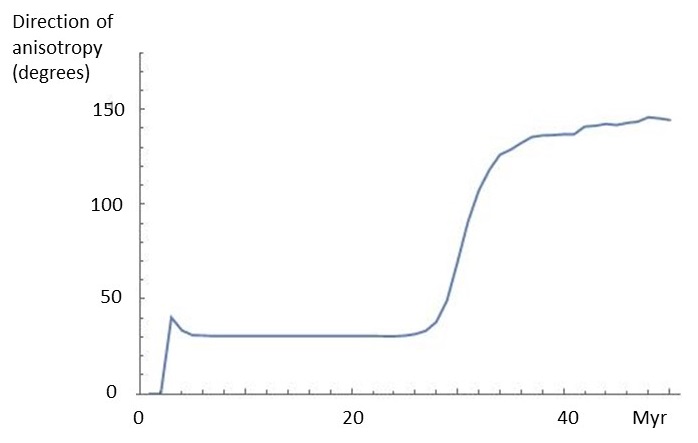}
\centering
\caption{
The angle in degrees of the direction of anisotropy at the Earth.  The angle is defined relative to the $z$ axis joining Cen A and M82
such that Cen A lies at angle $31^\circ$ and M82 at angle $149 ^\circ$.
}
\label{fig:figure1}
\end{figure}

\vskip .3 cm

\section{ ENERGETICS}

The UHECR fluxes from M82 and Cen A are roughly comparable at the present time.
Figure 2 therefore suggests that the flux from Cen A $\sim 20 {\rm Myr}$ ago needs to have been $\sim 200$ times greater than at present
if the flux of UHECR reflected from M82 is to have its present value.
In this section, we examine whether this is reasonable.
{
Estimates in this section are based on the assumption that (i) Cen A emits UHECR isotropically, and (ii) the TA and PAO hotspots are dominated by UHECR with an energy of 53EeV.
}

From figure 4 of di Matteo et al (2019) we make the very approximate estimate that UHECR with energies above 53EeV arrive at the Earth
from the M82 hotspot with an excess above the background at the rate of $\sim 5 \times 10^{-21} {\rm cm}^{-2} {\rm s}^{-1}$.
This estimate depends heavily on how a line is drawn around the complicated structure and fuzzy extent of the hotspot.
If the arrival rate directly from Cen A is 200 times greater than the arrival rate via M82, the UHECR luminosity of Cen A
20 Myr ago would have been approximately $1.4\times 10^{41} {\rm erg \ s}^{-1}$.
This is $2\times 10^{-5}$ of the Eddington luminosity of $7\times 10^{45}{\rm erg \ s}^{-1}$ for a Cen A black hole mass 
of $5.5 \times 10^7 {\rm M_\odot}$ (Cappellari et al 2009).
Assuming that 10 percent of the Cen A  Eddington luminosity was given to CR with energies above 1GeV, a CR spectral index of -2.3 between 1GeV and 50EeV would provide the required UHECR luminosity of the TA hotspot.
An efficiency of 10 percent and a spectral index of -2.3 is consistent with the efficiency and spectrum of CR production by high velocity shocks
in supernova remnants.

We therefore conclude that energy and power considerations are consistent with a model in which the M82 UHECR hotspot is an echo from Cen A.

\section{ IMPLICATIONS FOR COMPOSITION}

UHECR released into the IGM by Cen A are comprised of a range of energies, and probably also a range of charges $Ze$.
UHECR arriving at the Earth with the same energy may have different $Z$ and different scattering rates in the intergalactic medium, 
and hence different degrees of anisotropy.
An additional complication in our model is that 
UHECR arriving via M82 and directly from Cen A have a different history even if they were accelerated at the same time in Cen A.
UHECR with higher $Z$ but the same energy may be contained for longer within the lobes of Cen A in contrast to low $Z$ particles
that escape more easily.
Hence, UHECR arriving via M82 might be expected to have a lower $Z$ than those escaping later and arriving at the Earth directly from Cen A.

It would be consistent with the Matthews \& Taylor (2021) estimates of UHECR escape times from radio galaxies for 50EeV protons to escape in a few Myr but for higher $Z$ nuclei to leak out on the timescale of 20Myr.
If this applies to Cen A, the ratio of the flux from M82 to the direct flux fom Cen A may represent the relative production rates 20Myr ago of
light and intermediate mass UHECR nuclei.
It will be interesting to see whether this conjecture is confirmed by further data and refinements in the analysis of UHECR composition.

\section{ ECHOES FROM OTHER STARBURST GALAXIES}

We now add other strong nearby starburst galaxies to the model to explore their contribution to the UHECR sky.
From Table 3 of Ackermann et al (2012) we add NGC253 and IC342 at distances of 2.5 and 3.7Mpc respectively from Earth.
We do not include the strong starburst galaxy NGC4945 which is part of the Centaurus A group of galaxies and separated from
Cen A by a distance of only 480kpc (Tully et al 2015) which is comparable with the spatial extent of the lobes of Cen A.
UHECR reaching Earth from NGC4945 and Cen A are not separable either in angle on the sky or in their travel times to the Earth.
Any attempt to realistically model  CR propagation through NGC4945 would add considerable complexity to the
calculation without making much difference to the results.

The starburst galaxy M83 is also a member of the same group of galaxies as Cen A but at a greater distance from Cen A than NGC4945.
We tried including M83 in our model according to the same formalism adopted for NGC253 and IC342 and found that
it made little difference to the results.
For simplicity and because of the modelling incertainties we also omit M83 from the results presented here.

Diffusive reflections from more distant starburst galaxies contribute little in our model because of the  tendency towards a $r^{-4}$ 
dependence arising from the double action of the inverse square law for distant reflections.
This is consistent with the dominance of nearby starburst galaxies as shown in the analysis 
by  van Vliet et al (2021) of PAO anisotropies (Aab et al 2018, Caccianiga et al 2019).

NG253 and IC342 have radio luminosities at 1.4GHz that are smaller than that of M82 by factors of 0.39 and 0.35 respectively
(Ackermann et al 2012) and we assume that their UHECR reflectivities are smaller by the same ratios. 
We model this rather arbitrarily by reducing the areal cross-sections of their haloes.
The radii of their haloes are consequently taken to be $800\times\sqrt {0.39}$ and $800\times \sqrt{0.35}$ kpc respectively.
In other respects we model NGC253  and IC342 in the same way as M82.

Figure 5 is a plot against time of the UHECR flux from Cen A (blue line) and the combined flux at the Earth of UHECR reflected from 
the haloes of the three starburst galaxies (brown line). 
Comparison with Figure 2 gives a comparison of the fluxes from NGC253 and IC342
relative to that from M82.

\begin{figure}
\includegraphics[angle=0,width=8.5cm]{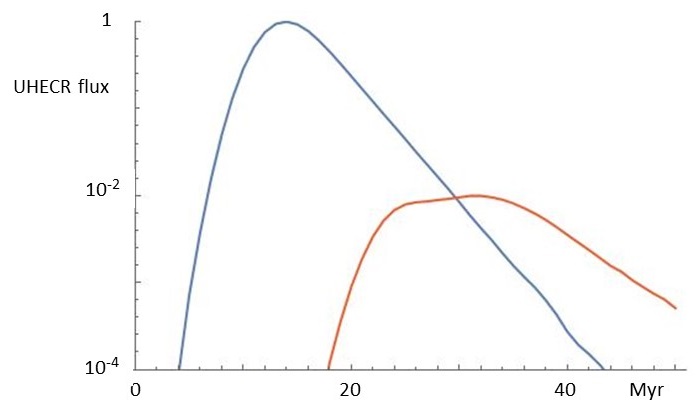}
\centering
\caption{
Time evolution of the logarithmic UHECR flux at the Earth (i)  UHECR arriving directly from Centaurus A without passing through
a starburst halo (blue line) (ii) UHECR arriving at the Earth having passed through the haloes of M82, NGC253 or IC342 (brown line).
The radius of the halo of M82 is 800kpc.
}
\label{fig:figure1}
\end{figure}


\begin{figure*}
\includegraphics[angle=0,width=\textwidth]{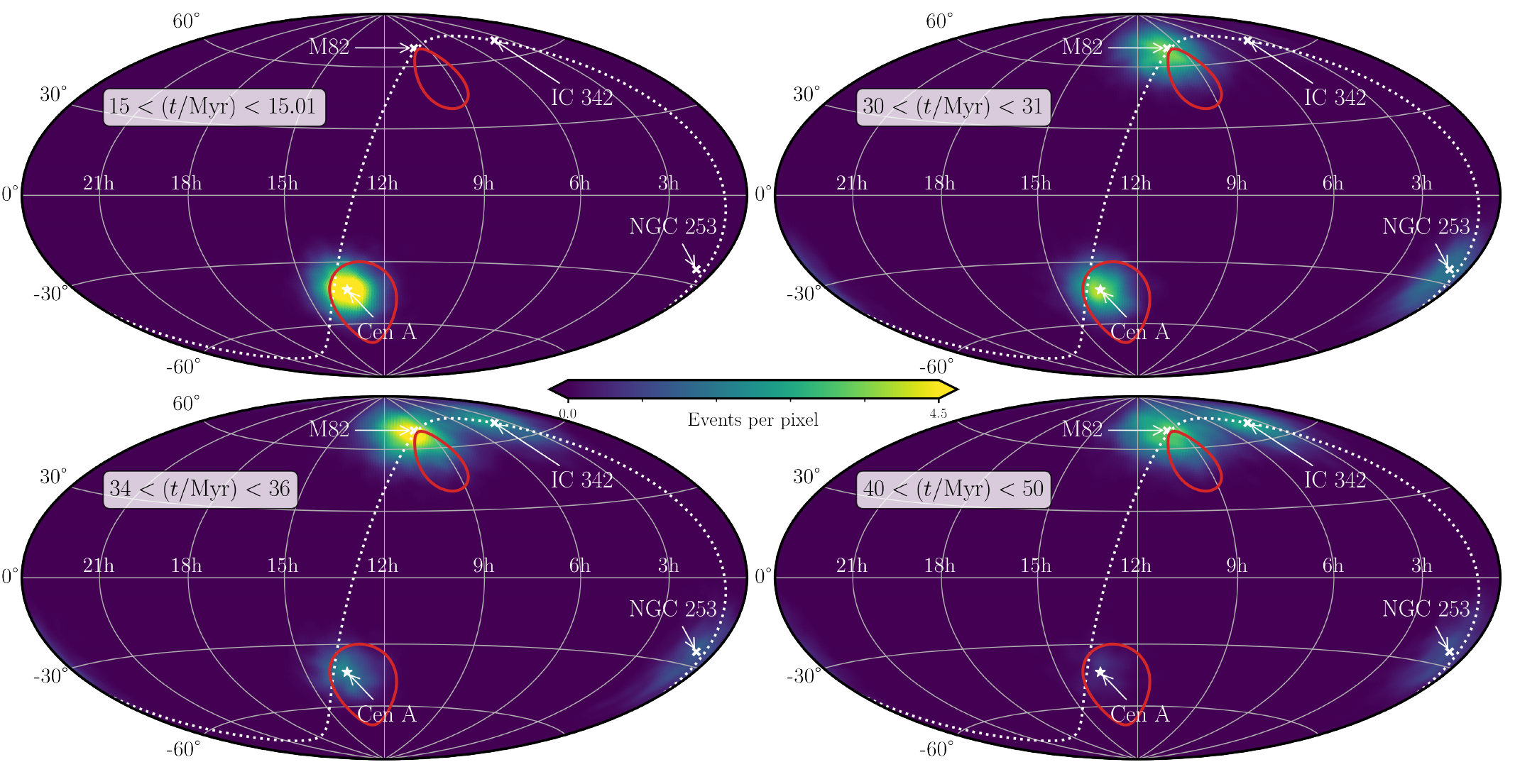}
\centering
\caption{
All-sky Hammer projection plot in equatorial coordinates (right ascension and declination) of arrival directions of Monte Carlo particles in four time bands of different duration: $15.0<t<15.01{\rm Myr}$ (Top Left), 
$30<t<31{\rm Myr}$ (Top Right), $34<t<36{\rm Myr}$ (Bottom Left), $40<t<50{\rm Myr}$ (Bottom Right). The three starburst galaxies included in the calculation (M82, NGC253 \& IC342) are labelled, as is Cen A, and the red lines mark the approximate positions of the TA and PAO excesses in the northern and southern hemispheres, respectively (Di Matteo et al. 2019). The plot is produced using Healpy, a python implementation of the HEALpix scheme (G{\'o}rski et al 2005; Zonca et al 2019). The colour-scale encodes the number of particles per HEALpix pixel, initially calculated with $64\times64$ pixels covering the sky, which has then been smoothed with a Gaussian symmetric beam with full-width at half-maximum of $5^\circ$. The colour-map can be thought of as a linear measure of UHECR intensity at Earth. The $34<t<36{\rm Myr}$ plot provides a reasonable qualitative match with the combined TA and PAO map from Di Matteo et al. (2019; see their figure 4). 
}
\label{fig:figure1}
\end{figure*}

The inclusion of NGC253 and IC342 disrupts the cylindrical symmetry of the calculation and we present the results as 
all-sky plots. Figure 6 plots the arrival directions of UHECR in equatorial coordinates in three time bands: (A) $15.0<t<15.01$Myr (Top Left),  close to the peak in the direct flux from Cen A
(B) $30<t<31$ Myr (Bottom Left), 
(C) $34<t<36$ Myr (Bottom Left), close to the peak in the flux from haloes 
(D) $40<t<50$Myr (Bottom Right),  
when the direct flux from Cen A has died down.
The choice of different time intervals scales the flux at the different times by the ratios 1:100:200:1000 respectively,
thus allowing the same colour scaling to be used on each plot.
The all-sky plot
uses the Hierarchical Equal Area isoLatitude Pixelization (HEALpix\footnote{http://healpix.sf.net}) scheme, so that Monte Carlo particles are binned according to which HEALpix pixel they fall into, with $64\times 64$ pixels covering the sky. 

Comparison with Figure 4 of di Matteo (2019) shows best agreement with the combined TA and PAO data in time band (C) at $t \approx 35{\rm Myr}$. 
The actual timing and the ratio between the direct and echoed fluxes depends on the escape time of 3Myr chosen in our calculation.
Nevertheless, figure 6 demonstrates that the Monte Carlo model is able to reproduce the TA and PAO fluxes when we choose our parameters suitably.

The TA and PAO data show UHECR arriving from a wider range of directions on the sky.
These may be reflections from more distant starburst galaxies of UHECR released by Cen A at earlier times.
Reflection may also occur from less prominent star-forming regions in multiple nearby galaxies.
Alternatively, UHECR may originate in other radio galaxies such as Fornax A (Matthews et al 2018) or a combination of multiple sources (Hardcastle et al 2010; Eichmann 2019a,b). 

\begin{figure}
\includegraphics[angle=0,width=8.5cm]{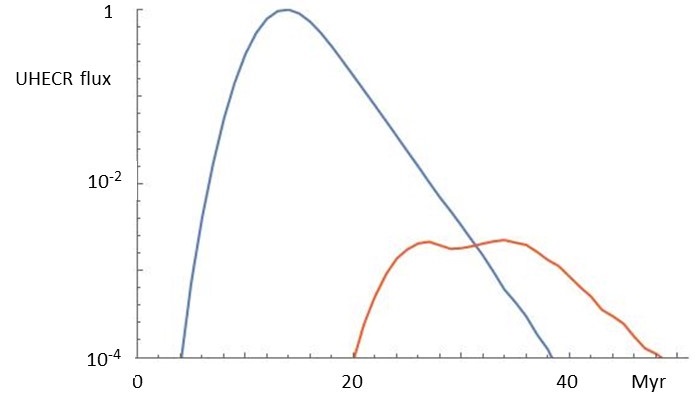}
\centering
\caption{
Time evolution of the logarithmic UHECR flux at the Earth (i)  UHECR arriving directly from Centaurus A without passing through
a starburst halo (blue line) (ii) UHECR arriving at the Earth having passed through the haloes of M82, NGC253 or IC342 (brown line).
The radius of the halo of M82 is 400kpc in contrast to figure 5 where the radius was 800kpc.
}
\label{fig:figure1}
\end{figure}

\begin{figure*}
\includegraphics[angle=0,width=\textwidth]{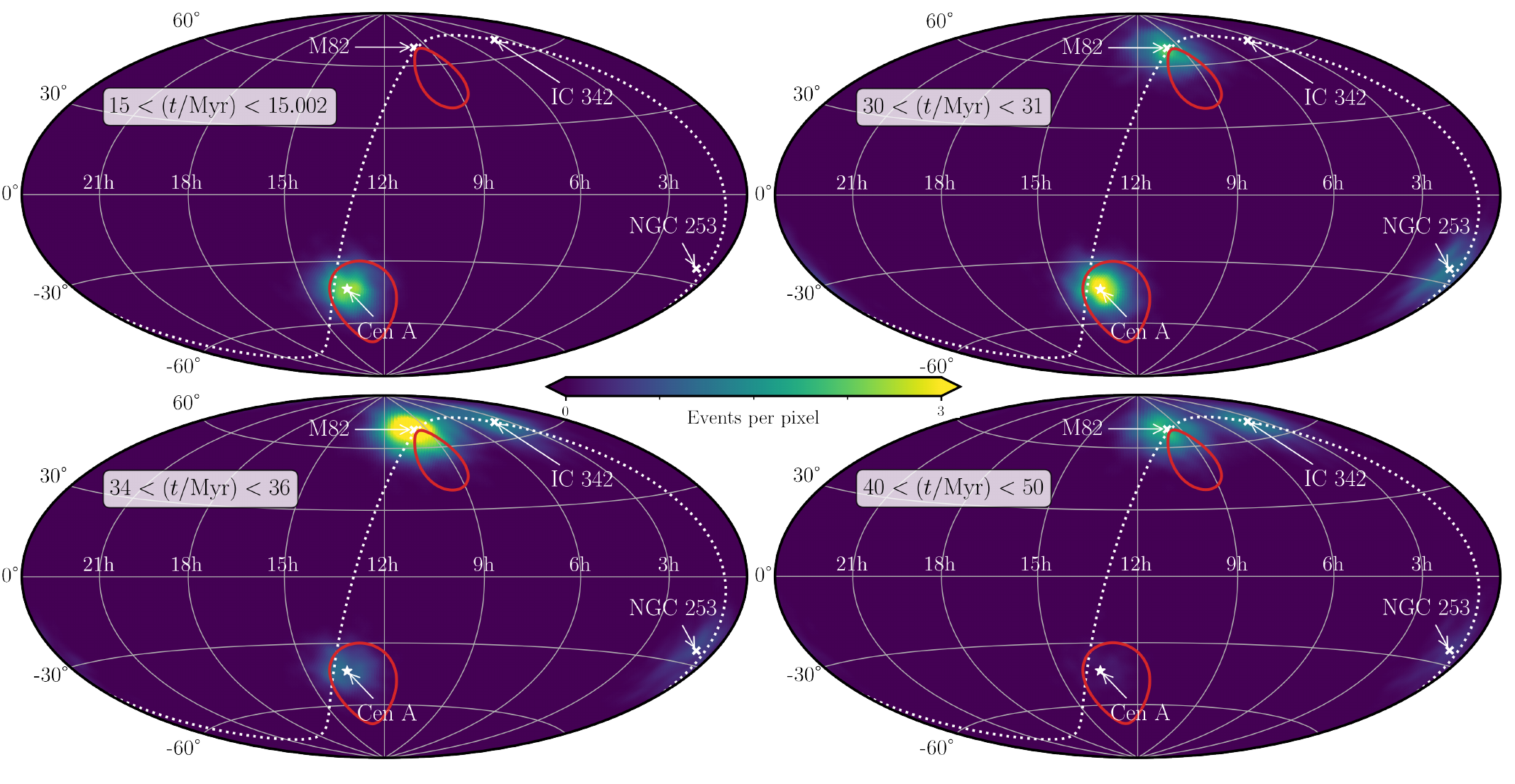}
\centering
\caption{
The same as for figure 6 except that the halo radii of M82 and the other starburst galaxies are decreased by a factor of two, the scattering rate in the haloes is doubled,
and the UHECR release time from Cen A 
is reduced from 3Myr to 2.5Myr.
The first time band  ($15.0<t<15.002 $) is five times shorter than in figure 6 in order to compensate for the relatively larger UHECR luminosity
of Cen A when halo radii are reduced.
}
\label{fig:figure1}
\end{figure*}

\section{ A SMALLER HALO}

As discussed in Section 2, we now repeat the above calculations with a halo radius of 400kpc instead of 800kpc.
The same halving of the halo radius is applied to NGC253 and IC342.
The magnetic field can be expected to decrease with distance from the galaxy as in equation (2) so we
make the scattering rate for the 400kpc halo twice that for the 800kpc halo considered above.
This maintains the same optical depth of the halo to UHECR propagation.

We also change the decay time of the UHECR release from Cen A from 3Myr to 2.5Myr so that the ratio of the fluxes from Cen A and M82 are in 
a similar proportion at 35Myr as they were for 800kpc radius.
This is apparent in figure 7 where the flux from M82 is reduced by a factor of four because the cross-section presented by the halo to UHECR
propagation is reduced in proportion to the square of its radius.

The all-sky Hammer plot for the smaller radius is presented in figure 8 in four time bands of different duration: $15.0<t<15.002 {\rm Myr}$ (Top Left), 
$30<t<31{\rm Myr}$ (Top Right), $34<t<36{\rm Myr}$ (Bottom Left), $40<t<50{\rm Myr}$ (Bottom Right).
The second,  third and fourth time bands are the same as in figure 6.
The first time band ($15.0<t<15.002 $) is 5 times shorter than in figure 6 in order to keep the number of UHECR arrivals in each time band comparable.

Figure 8 appears similar to figure 6 for which the halo radius was 800kpc.
The main effect of reducing the halo radius of M82 is that the UHECR power of Cen A 20Myr ago has to be made four times larger to maintain the same flux
at the Earth
from UHECR scattered by M82.
With this compensation,
and referring back to the energetics calculation in Section 4, the UHECR luminosity of Cen A 20Myr ago would need to have been
$5.6\times 10^{41} {\rm erg \ s}^{-1}$ instead of $1.4\times 10^{41} {\rm erg \ s}^{-1}$.
A straightforward multiplication of the total CR power of Cen A would make it 40\% instead of 10\% of the Eddington luminosity.
While not unreasonable, an alternative is to keep the CR luminosity at 10\% of the Eddington luminosity
and assume that the CR energy spectral index is 2.25 instead of 2.3,
thus increasing the number ratio of CR at 50EeV to CR at 1GeV by a factor of four
while keeping unchanged the total CR power.

\section{ CONCLUSIONS}

In summary, we have shown that a viable model can be constructed in which the TA hotspot is an echo of earlier UHECR production by Cen A during 
a period of greater activity 20Myr ago.  
The suggested echo is due to enhanced UHECR scattering by the M81 group of galaxies and 
in particular by the halo of the starburst galaxy M82.
Echoes from other starburst galaxies or other regions of enhanced magnetic field may be responsible for UHECR arriving from other parts of the sky.
The model has the potential to reconcile the statistical association of UHECR with starburst galaxies and the inability,
due to their relatively low power, 
of starburst galaxies themselves to generate UHECR of sufficient ultra-high energy.
Differences in UHECR composition may result from the varied histories of UHECR arriving at the Earth by different paths.
As more data is collected, UHECR may be used to probe the local intergalactic environment and the structure of starburst haloes out to distances of 10-20 Mpc from the Earth.
The magnitude and structure of the magnetic field in the outer parts of starburst haloes 
are a major uncertainty in the model.

\section*{Acknowledgments}
We thank Andrew Taylor and Sergio Martin-Alvarez for helpful discussions. 
We acknowledge the contribution made by the referees to the improvement of this paper.
ARB acknowledges the support of an Emeritus Fellowship from the Leverhulme Trust.
JM acknowledges a Herchel Smith Fellowship at the University of Cambridge. Some of the results in this paper have been derived using the healpy and HEALPix packages (G{\'o}rski et al 2005; Zonca et al 2019). We also gratefully acknowledge the use of Astropy, a community-developed core Python package for astronomy (Astropy Collaboration 2013, 2018), and Matplotlib v3.1.1 (Hunter 2007). 
The Monte Carlo calculations were performed using the computing resources provided by STFC Scientific Computing Department’s SCARF cluster.

\section*{Data Availability}
The data underlying this article are available from the authors on reasonable request.

\section*{References}
Aab A et al (Pierre Auger collaboration) 2018 ApJ 853, L29
\newline
Abbasi RU et al (The Telescope Array Collaboration) 2018  ApJ 867, L27
\newline
Ackermann M e al 2012 ApJ 755, 164
\newline
{
Adebahr B, Krause M, Klein U, Heald G, Dettmar R-J 2017 A\&A 608, 29
\newline
Adebahr B, Krause M, Klein U, Wezgowiec M, Bomans DJ, Dettmar R-J 2013 A\&A  555, 23
}
\newline
Aharonian FA, Belyanin AA, Derishev EV, Kocharovsky VV, Kocharovsky VlV 2002 Phys Rev D 66, 023005
\newline
{
Alves Batista R., Dundovic A., Erdmann M., Kampert K.-H., Kuempel D., M{\"u}ller G., Sigl G., et al., 2016, JCAP, 2016, 038
}
\newline
Alves Batista et al 2019 Frontiers Astr Space Sci 6, 23 
\newline
Anchordoqui LA 2017, Proc European Physical Society Conf. on High
Energy Physics (EPS-HEP2017), Venice, Italy. p. 1, 
http://adsabs.harvard.edu/abs/2017ehep.confE...1A
\newline
Anchordoqui LA    2018   Phys Rev D 97, 063010
\newline
Astropy Collaboration, 2013, A\&A, 558, A33.
\newline
Astropy Collaboration, 2018, AJ, 156, 123.
\newline
Baerwald P, Bustamente M, Winter W 2015 Astropart Phys 62, 66
\newline
{
Becker JK, Stamatiko M, Halzen F, Rhode W 2006 Astropart Phys 25, 118
}
\newline
Beckmann RS et al 2019 A\&A 631, A60
\newline
Bell AR, Araudo AT, Matthews JH, Blundell KM 2018 MNRAS 473, 2364
\newline
Bertone S, Vogt C, Enßlin T 2006 MNRAS 370, 319
\newline
{
Biermann PL, de Souza V  2012 ApJ 746, 72
}
\newline
Birdsall CK, Langdon AB 1985 {\it Plasma Physics via Computer Simulation}, publ McGraw-Hill, New York.
\newline
Bister T et al, (Pierre Auger Collaboration) 2021 Phys Scripta 96, 074003
\newline
Blandford RD 2000 Physica Scripta T85, 191
\newline
Boncioli D, Biehl D, Winter W 2019 ApJ 872, 110
\newline
Borthakur S, Heckman T, Strickland D, Wild V, Schiminovich D 2013 ApJ 768:18
\newline
Caccianiga L for the Pierre Auger Collaboration 2019 ICRC2019, 206,  
https://pos.sissa.it/cgi-bin/reader/conf.cgi?confid=358, id.206
\newline 
Cappellari M, Neumayer N, Reunanen J, van der Werf P, de Zeeuw PT, Rix H-W 2009 MNRAS 394, 660
\newline
Cavagnolo KW, McNamara BR, Nulksen PEJ, Caroilla CL, Jones C, Birzan L 2010 ApJ 720, 1066
\newline
de Cea del Pozo E, Torres DF, Marrero AYR 2009 ApJ 698, 1054
\newline
Dermer CD 2011 AIPC 1358, 355
\newline
de Mello DF, Smith LJ, Sabbi E, Gallagher JS, Mountain M, Harbeck DR 2008 AJ 135, 548
\newline
di Matteo A et al 2019 Proc International Cosmic Ray Conf (ICRC) 36, 439
\newline
Domingo-Santamaria E, Torres DF 2005 A\&A 444, 403
\newline
Donnert J, Vazza F, Brüggen M, ZuHone J 2018 Space Sci Rev 214, 122
\newline
Eichmann B, Rachen JP, Merten L, van Vliet A, Becker Tjus J 2018 JCAP02(2018)036
\newline 
Eichmann B 2019a JCAP05(2019)009
\newline
Eichmann B 2019b Proc International Cosmic Ray Conf (ICRC) 36, 245
\newline
Eilek JA 2014 New J Phys 16, 045001
\newline
{
 Ferri\`ere K, Terral P 2014 A\&A 561 1000
}
\newline
Gaspari M, Temi P, Brighenti, F 2017 MNRAS 466, 677
\newline
G{\'o}rski KM, Hivon E, Banday AJ, Wandelt BD, Hansen FK, Reinecke M, Bartelmann M, 2005, ApJ 622, 759
\newline
Greco JP, Martini P, Thompson TA 2012 ApJ 757:24
\newline
Greisen K 1966 Phys Rev Lett 16, 748
\newline
Guedes Lang R, Taylor AM, Ahlers M, de Souza V, 2020, Phys.Rev.D 102, 063012
\newline
{
Guetta D, Hooper D, Alvarez-Mu{\~n}iz J, Halzen F, Reuveni E 2004 Astropart Phys 20, 429
}
\newline
Hardcastle MJ, Cheung CC, Feain IJ, Stawar L, 2009 MNRAS 393, 1041
\newline 
Hardcastle MJ, 2010, MNRAS, 405, 2810
\newline
Heckman TM, Armus L, Miley GK 1990 ApJS 74, 833
\newline
Heinze J, Biehl D, Fedynitch A, Boncioli D, Rudolph A, Winter W 2020 MNRAS 498, 5990
\newline
Hillas AM 1984 ARAA 22, 425
\newline
Hooper D, Sarkar S, Taylor AM 2007 Astropart Phys 27, 199
\newline
Hunter JD 2007, Comput. Sci. Eng., 9, 90
\newline
{
Irwin J, Damas-Segovia A, Krause M, Miskolczi A, Jiangtao Li, Stein Y, English J,
Henriksen R, Beck R, Wiegert T, Dettmar R-J
2019 Galaxies 7, 42
}
\newline
Joshi JC, Mirand LS, Razzaque S, Yang L 2018 MNRAS 478, L1
\newline
Karachentsev ID 2005 Astronomical Journal 129, 178
\newline
Karachentsev ID, Kashibadze OG 2006 Astrophysics 49, 3
\newline 
Kawata K et al 2015 Proc ICRC2015, 276
\newline
Lemoine M, Pelletier G 2010 MNRAS 402, 321
\newline
Lopez-Rodriguez E, Guerra JA, Asgari-Targhi M, Schmel JT  2021  ApJ 914, 24
\newline 
Lovelace RVE 1976 Nature 262, 649
\newline 
Mackie G, Fabbiano G 1998 AJ 115, 514
\newline 
Maccagni FM, Murgia M, Serra P, Govoni F, Morokuma-Matsui K, Kleiner D, Buchner S et al 2020, A\&A, 634, A9
\newline
Massaglia S 2007 Nucl Phys B Proc Suppl 165, 130
\newline
Matthews JH, Bell AR, Blundell KM, Araudo AT 2018 MNRAS 479, L76
\newline
Matthews JH, Bell AR, Blundell KM, Araudo AT 2019 MNRAS 482, 4303
\newline
Matthews JH, Bell AR, Blundell KM 2020  New Astronomy Reviews, 89, 101543
\newline
Matthews JH, Taylor AM 2021 MNRAS 503, 5948
\newline
{
Merten L, Becker Tjuis J, Fichter H, Eichmann B, Gunter S 2017 JCAP 6, 46
}
\newline
O’Sullivan S, Reville B, Taylor AM 2009 MNRAS 400, 24
\newline
Persic M, Rephaeli Y 2020 MNRAS 491, 5740
\newline
Ptitsyna V, Troitsky SV 2010 Phys Usp 53, 691
\newline
Reville B, Bell AR 2014 MNRAS 439, 2050
\newline
Romero GE, Muller AL, Roth M  2018  A\&A 616, A57
\newline
Rudolph A, Bosnjak Z, Palladino A, Sadeh I, Winter W 2021 arXiv:210704612
\newline 
Russell HR, McNamara BR, Edge AC, Hogan MT, Main RA, Vantyghem AN 2013 MNRAS 432, 530
\newline
Samuelsson F, Begue D, Ryde, Pe'er A 2019 ApJ 876, 93
\newline
Samui S, Subramanian K, Srianand R 2018 MNRAS 476, 1680
\newline
Smercina A, Bell EF, Price PA, Slater CT, D'Souza R, Bailin J, de Jong RS, Jang IS, Manochesi A, Nidever D 2020
ApJ 905, 60
\newline
Stecker FW, Salamon MH 1999 ApJ 512, 521
\newline
Strickland DH, Heckman TM 2009 ApJ 697, 2030
\newline
Strickland DH, Stevens IR 2000 MNRAS 314, 511
\newline
{
Todero Peixoto C et al (Pierre Auger collaboration) 2019   
Proc International Cosmic Ray Conf (ICRC) 36, 440
}
\newline
Tully BR, Libeskind NI, Karachentsev IG, Karachentseva VE, Rizzi L, Shaya EL 2015 ApJ 802 L25
\newline
Tumlinson J, Thom C, Werk JK, Prochaska JX, Tripp TM, Weinberg DH, Peeples MS, O'Meara JM, Oppenheimer BD,  Meiring JD,  Katz NS,  Davé R, Ford AB,
 Sembach KR 2011 Science 334, 948
\newline
van Velzen S, Falcke H, Schellart P, Nierstenhofer N, Kamper K-H 2012 A\&A 544, A18
\newline
van Vliet A, Palladino A, Taylor A, Winter W 2021
arXiv:2104.05732
\newline
{Vietri  M 1995 ApJ 453, 883}
\newline
Wang J, Hammer F, Rejbuka M, Crnojevic D, Yang Y 2020 MNRAS 498, 2766
\newline
Waxman E 1995 PRL 75, 386
\newline
Waxman E, Bahcall JN 1997 Phys Rev Lett 78, 2292
\newline
Waxman E, Bahcall JN 1998 Phys Rev D 59, 023002
\newline
Waxman E, 2001 In Physics and Astrophysics of Ultra-High-Energy Cosmic Rays, Edited 
by M. Lemoine, G. Sigl, Lecture Notes in Physics, vol. 576, p.122, eprint: arXiv:astro-
ph/0103186. 
\newline
Wilde MC et al 2021 ApJ 912:9
\newline
Wykes et al 2013 A\&A 558, A19
\newline
Yang H-Y K, Reynolds CS 2016 ApJ 829, 90
\newline
Zatsepin GT, Kuz’min VA 1966 Sov J Expt Theor Phys Lette 4, 78
\newline
Zhang BT, Murase K, Kimura SS, Horiuchi S, Meszaros P 2018 Phys Rev D 97, 083010
\newline
Zonca A, Singer L, Lenz D, Reinecke M, Rosset C, Hivon E, G{\'o}rski KM 2019 The Open Journal 4, 1298

\end{document}